# Substitutional oxygen as the origin of the 3.5 eV luminescence in hexagonal boron nitride

*Marek Maciaszek[1,2]\**

1 - Faculty of Physics, Warsaw University of Technology, Koszykowa 75, 00-662 Warsaw, Poland

2 - Center for Physical Sciences and Technology (FTMC), Vilnius LT-10257, Lithuania

*Abstract:* Although point defects in hexagonal boron nitride exhibiting single-photon emission attract considerable interest, a broader understanding of defect physics and chemistry in hBN remains limited, potentially hindering further development. Oxygen is among the most common impurities in hBN, and numerous studies have reported a pronounced photoluminescence band centered near 3.5 eV following oxygen incorporation, yet its microscopic origin has remained unresolved. Here, we demonstrate that this emission originates from hole capture by neutral oxygen substituting for nitrogen ($O_N$). The transition mechanism is non-trivial, involving not only a change in charge state but also a substantial structural reconfiguration: the positive and neutral states exhibit markedly different geometries and symmetries. In the neutral state the defect adopts a low-symmetry configuration with out-of-plane displacements of the oxygen and neighboring atoms. The calculated emission energy (3.63 eV) and lineshape are in excellent agreement with experiment.

---

\* Email: marek.maciaszek@pw.edu.pl

In recent years, hexagonal boron nitride (hBN) has emerged as one of the most promising materials for applications in quantum technologies. This interest was sparked by the discovery of single-photon emitters (SPEs) in hBN exhibiting outstanding optical properties, including high brightness even at room temperature, excellent photostability, and narrow emission linewidths [1, 2]. Moreover, the layered nature of hBN enables straightforward integration of these emitters with photonic waveguides and optical cavities, making the material particularly attractive for on-chip quantum photonics [3].

The observed single-photon emission is associated with point defects in the hBN lattice, whose microscopic structure and origin remain the subject of extensive debate [4]. While some defects are responsible for single-photon emission, many others are optically inactive or give rise to different luminescence features [5, 6]. Nevertheless, a comprehensive understanding of the physics and chemistry of point defects in hBN is essential. Such knowledge is crucial not only for identifying the true microscopic origin of SPEs, but also for optimizing material growth, improving sample quality, and enabling tailoring of hBN for specific quantum and optoelectronic applications. Further development of hBN-based technologies therefore critically depends on a deeper, defect-level understanding of this material.

Oxygen is believed to be one of the most common impurities in hexagonal boron nitride (hBN) [7]. Its presence has been experimentally confirmed using secondary-ion mass spectrometry (SIMS) [8], X-ray photoelectron spectroscopy (XPS) [9] as well as annular dark-field scanning transmission electron microscopy (ADF-STEM) [10]. Among oxygen-related defects, $O_N$, $V_BO_N$, and $C_NO_N$ are predicted to be the most abundant [11, 12]. One of these defects, $V_BO_N$, has been proposed as a spin center with a zero-phonon line (ZPL) at 2.1 eV [13]. The impact of oxygen on the photoluminescence (PL) spectra of hBN has been investigated in several studies [14-16]. Although some basic properties of oxygen-related defects have been explored theoretically [11, 17], a more detailed description of defect-lattice coupling is still lacking, yet such an analysis would be beneficial for a complete interpretation of experimentally observed PL features.

Experimentally, the incorporation of oxygen into hBN has been shown to induce or significantly enhance a broad emission band centered around 3.5 eV [14-16]. This emission exhibits a large linewidth, with a full width at half maximum (FWHM) of approximately 0.7 eV [15]. In reported studies, oxygen was introduced using different approaches, including in situ oxygen doping during hydride vapor phase epitaxy (HVPE) growth [15] and post-growth oxygen plasma treatment [16]. Despite differences in the incorporation approach, this emission feature appears systematically across studies and is notably intense; however, its microscopic origin remains unresolved.

In this study, we demonstrate that hole capture by substitutional oxygen defect on the nitrogen site ($O_N$) is responsible for the PL band centered at 3.5 eV observed in oxygen-containing hBN samples [14-16]. Although the charge-state transition level of $O_N$ lies high in the bandgap (5.09 eV above the valence-band maximum, VBM), the resulting emission is strongly red-shifted due to pronounced lattice relaxation accompanying hole capture by the neutral defect. In particular, the neutral charge state stabilizes in a low-symmetry equilibrium configuration, characterized by out-of-plane displacements of the oxygen atom and one of its neighboring boron atoms. Upon hole capture, the defect relaxes into a high-symmetry configuration with all atoms occupying in-plane sites, corresponding to the positive charge state. This substantial structural reconfiguration releases a large relaxation energy, thereby lowering the peak energy of the optical transition to 3.63 eV. The calculated emission lineshape exhibits a full width at half maximum (FWHM) of 0.69 eV, in good agreement with the experimental value of approximately 0.7 eV.

Ab initio calculations were performed using spin-polarized density functional theory (DFT) as implemented in VASP [18], employing the projector-augmented wave (PAW) method [19] with a plane-wave energy cutoff of 500 eV. The exchange–correlation energy was described using the hybrid HSE functional [20], with the mixing parameter increased to 0.31 to reproduce the experimental band gap. The calculations were carried out using a 288-atom supercell (6x6x2), and the Brillouin zone was sampled at the Γ point only. Atomic positions were relaxed until the forces acting on all ions were below 0.01 eV/Å (and 0.005 eV/Å for phonon frequency calculations). To account for van der Waals interactions, the Grimme D3 dispersion correction was applied [21]. This methodology yields a band gap of 5.95 eV, lattice constants of a=2.490 Å and c=6.558 Å, and a compound formation enthalpy $\Delta H_f$(hBN) of -2.89 eV. These values are in good agreement with experimental data (a=2.506 Å, c=6.603 Å, and $\Delta H_f$(hBN)=-2.60 eV) reported in [22, 23].

The discussion begins with an analysis of the formation energy and geometries of the $O_N$ defect in the charge states found to be stable. The formation energy of the $O_N$ defect was calculated according to the formula [24, 25]:

$$E^f(q) = E_D(q) - E_H - \mu_O + \mu_N + q(E_V + E_F) + E_{corr}$$

where $E_D(q)$ is the total energy of the supercell containing a single $O_N$ defect in charge state q, $E_H$ is the total energy of the defect-free supercell, $\mu_O$ and $\mu_N$ are the chemical potentials of oxygen and nitrogen, respectively, $E_F$ is the Fermi level referenced to the VBM ($E_V$), and $E_{corr}$ is the correction term accounting for spurious electrostatic interactions between periodically repeated charged supercells. $E_{corr}$ was evaluated using the Freysoldt-Neugebauer-Van de Walle scheme [26]. The charge-state transition level $\varepsilon(q/q')$ is defined as the Fermi level at which the formation energies of the two charge states, q and q', are equal.

Two limiting growth conditions were considered: N-rich and N-poor. Under N-rich conditions, the chemical potential of nitrogen is maximal and is determined by the nitrogen molecule: $\mu_N = \frac{1}{2}E(N_2)$, while the chemical potential of boron reaches its minimum value, $\mu_B = \mu_B^{solid} + \Delta H_f(hBN)$, where $\mu_B^{solid}$ is the chemical potential of boron in its bulk crystalline phase. Under N-poor conditions, $\mu_B = \mu_B^{solid}$, and $\mu_N = \frac{1}{2}E(N_2) + \Delta H_f(hBN)$. Because of the high thermodynamic stability of $B_2O_3$ (our calculated value of $\Delta H_f(B_2O_3)$ is -12.88 eV), the oxygen chemical potential is constrained by the formation enthalpy of $B_2O_3$. Consequently, under N-rich conditions is $\mu_O = \frac{1}{3}[E(B_2O_3) - 2\mu_B^{solid} - 2\Delta H_f(hBN)]$, while under N-poor conditions is $\mu_O = \frac{1}{3}[E(B_2O_3) - 2\mu_B^{solid}]$. We note that the thermodynamic stability of $B_2O_3$ imposes a stricter bound on $\mu_O$ than $O_2$.

Two charge states of the $O_N$ defect were found to be stable: positive (q=+1) and neutral (q=0). For the positive charge state, all atoms occupy in-plane positions, and the defect configuration exhibits a high symmetry ($D_{3h}$). The calculated formation energy in this charge state is in good agreement with previous results [11]. In contrast, for the neutral charge state, the symmetry of the defect configuration is significantly reduced (only $C_1$). In this low-symmetry configuration, the oxygen atom and its neighboring boron atoms undergo out-of-plane displacements. The oxygen atom is displaced by 0.136 Å from the layer plane; two boron atoms are displaced by 0.070 Å on the same side of the layer as oxygen, while the third boron atom is displaced by 0.368 Å to the opposite side of the layer. Figure 1 shows the relaxed geometries of both charge states, together with the corresponding bond lengths.

For the neutral charge state, the high-symmetry defect configuration with all atoms in-plane is $E_{HS-LS}$=0.33 eV higher in energy than the low-symmetry configuration. Nudged elastic band (NEB) calculations [27, 28] reveal no or very low energy barrier for relaxation from the high-symmetry to the low-symmetry geometry indicating that the distortion occurs spontaneously (Fig. 2b). Since the localized defect level lies close to the conduction band minimum, the NEB calculations were performed using the HSE functional, consistent with the rest of the calculations.

The (1+/0) charge-state transition level is located at 5.09 eV above the VBM. Since the equilibrium Fermi level in hBN is expected to lie below this level under most conditions [11, 12], the $O_N$ defect is predicted to exist predominantly in the positive charge state in thermodynamic equilibrium. If the high-symmetry geometry is imposed for the neutral charge state, the transition level shifts to $\varepsilon(+/0) + E_{HS-LS} = 5.42$ eV above the VBM, bringing it into agreement with the value reported in [11]. The formation energy of $O_N$ under N-rich and N-poor conditions is shown in Fig. 2a; results corresponding to the unstable high-symmetry geometry of the neutral charge state are also included for comparison (dashed lines).

We now analyze the optical properties of the $O_N$ defect. Since this defect introduces only a single localized state into to the bandgap, its optical behavior is primarily governed by interactions with free carriers. The equilibrium charge state of $O_N$ is +1; therefore, under illumination, when free carriers are present, the defect can capture an electron and transition to the neutral charge state. Assuming that the energy barrier for structural relaxation is negligible or sufficiently low to be thermally activated, the defect subsequently relaxes from the high-symmetry geometry associated with the positive charge state to the low-symmetry geometry associated with the neutral state.

In the presence of holes, $O_N$ in the neutral charge state can then capture a hole and return to the +1 charge state. Because the geometries of the initial (neutral) and final (positive) charge states differ substantially, this transition is accompanied by a large structural rearrangement, resulting in strong electron-phonon coupling. The calculated vertical emission energy is $E_{em}$=3.63 eV, and the associated relaxation energy (Franck-Condon shift) is $E_{rel}$ is 1.46 eV. In the regime of strong electron-phonon coupling, the vertical emission energy corresponds to the maximum of the PL spectrum [29]. The corresponding configuration coordinate diagram shows Fig. 3.

To determine the luminescence lineshape, we begin with a quantitative analysis of the electron–phonon coupling. Using the effective-mode approximation [30, 31], the average phonon energy $\hbar\omega_{av}$ is calculated to be 59.6 meV, and the resulting Huang–Rhys factor $S$ is 24.4. This large value places the system in the strong electron–phonon coupling regime, under which the one-dimensional configuration-coordinate model provides an excellent approximation to the luminescence lineshape [31].

Within the 1D model, the luminescence lineshape is described by

$$I(E) \propto \sum_n e^{-S} \frac{S^n}{n!} g_\sigma(E_{ZPL} - n\hbar\omega_{av} - E)$$

where $g_\sigma$ is a Gaussian function with a smearing parameter $\sigma$. For the transition considered here, the zero-phonon-line energy $E_{ZPL}$ is determined by the (+/0) charge-state transition level and is equal to 5.09 eV. The calculated luminescence lineshape is shown in Fig. 4. For comparison, experimental spectra of undoped and oxygen-doped hBN samples from [15] are also shown, demonstrating an increase in the intensity of the 3.5 eV band following oxygen incorporation. To facilitate a direct comparison of the lineshapes, the calculated spectrum was rigidly red-shifted by 0.10 eV. The FWHM of the calculated spectrum is 0.69 eV. The calculated lineshape is in very good agreement with the experimental peak 3.5 eV.

The experimental spectrum reported in [15] was obtained under above-bandgap illumination (195 nm), which naturally leads to the generation of free carriers. Consequently, hole capture by defects is expected. In contrast, [16] employed sub-band gap excitation (325 nm). However, even under sub-bandgap illumination, free carriers may be generated in the presence of other defects through defect-mediated processes. For example, it has been proposed that the boron vacancy can generate both electrons and holes via a three-step mechanism involving internal excitation, photoionization, and optical hole emission, with a threshold energy of 2.45 eV [32]. Similar processes may operate for other defects, making the presence of free carriers under sub-bandgap excitation a plausible assumption [33, 34]. Additionally, it must be noted that the oxygen-plasma treatment employed in study [16] leads to an apparent upward shift of the VBM and a reconstruction of the band-edge density of states, which may influence the spectral position of the PL peak in such samples and render a direct comparison between calculations and experiment for this class of samples less straightforward.

An analysis of the experimental results in [15] and [16] shows that oxygen incorporation enhances not only the 3.5 eV peak but also the peaks at 5.33 eV [15] or 2.3 and 2.5 eV [16]. The addition of oxygen likely passivates nitrogen vacancies leading to an overall improvement in sample quality. As a result, PL from bound excitons is expected to be stronger, which may explain the enhanced emission around 5.3 eV, a spectral region where excitonic transitions are typically observed [6, 35].

Moreover, many hBN samples contain a significant concentration of boron vacancies [5, 36]. Owing to electrostatic attraction, these vacancies, typically negatively charged, can bind to the $O_N$ donor to form stable neutral $V_BO_N$ complexes. The binding energy of such a complex can be estimated to be approximately 3.6 eV [11]. This process is feasible at temperatures above ~900 K, where boron vacancies become mobile [37]. The relatively rich electronic structure of the $V_BO_N$ complex [13] may explain the appearance of additional transitions (e.g., at 2.3 and 2.5 eV) in the PL spectrum.

In summary, we demonstrate that $O_N$ is a strong candidate for the origin of the 3.5 eV PL peak in hBN. The calculated vertical emission energy (3.63 eV) and the FWHM (0.69 eV) of the peak associated with hole capture by neutral $O_N$ are in very good agreement with experimental data for oxygen-doped hBN samples.

ACKNOWLEDGMENTS

Computational resources were provided by the Interdisciplinary Center for Mathematical and Computational Modelling (ICM), University of Warsaw (Grant No. GB81-6), and by the High Performance Computing Center "HPC Saulėtekis" in the Faculty of Physics, Vilnius University, Lithuania.

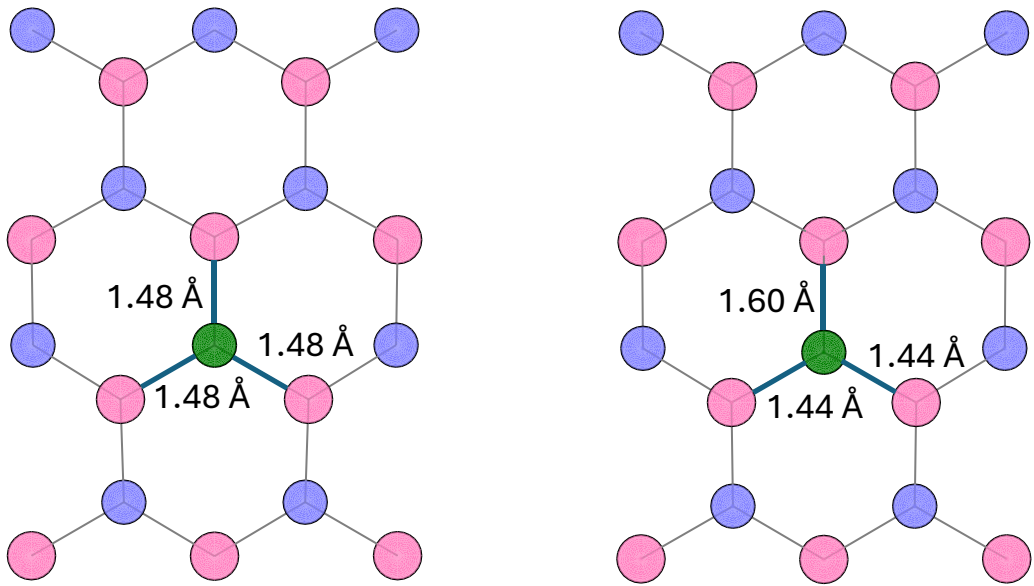

Fig. 1. Equilibrium geometries of the O$_N$ defect in the positive (left) and neutral (right) charge states. Boron, nitrogen, and oxygen atoms are represented by pink, blue, and green spheres, respectively. Bond lengths are provided.

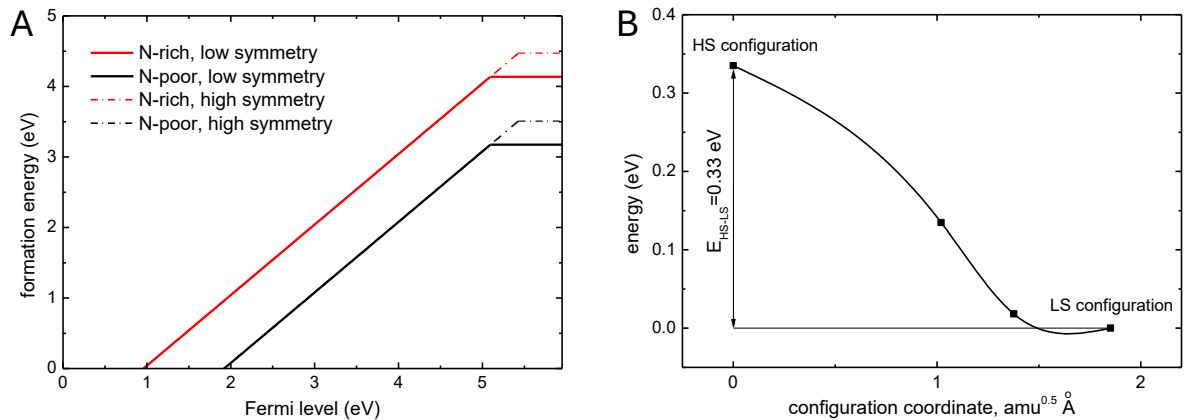

Fig. 2. (a) Formation energy of the $O_N$ defect under N-rich (red line) and N-poor (black line) conditions. Solid lines indicate the results corresponding to the equilibrium geometries of both stable charge states. Dashed lines indicate the formation energies assuming that the high-symmetry configuration is the equilibrium geometry of the neutral charge state. (b) Calculated potential energy surface for the transition of neutral $O_N$ from the high-symmetry (HS) configuration, with all atoms in-plane, to the low-symmetry configuration in which the O atom and three neighboring B atoms occupy out-of-plane sites. Squares represent results of the NEB calculation and are connected by spline curves.

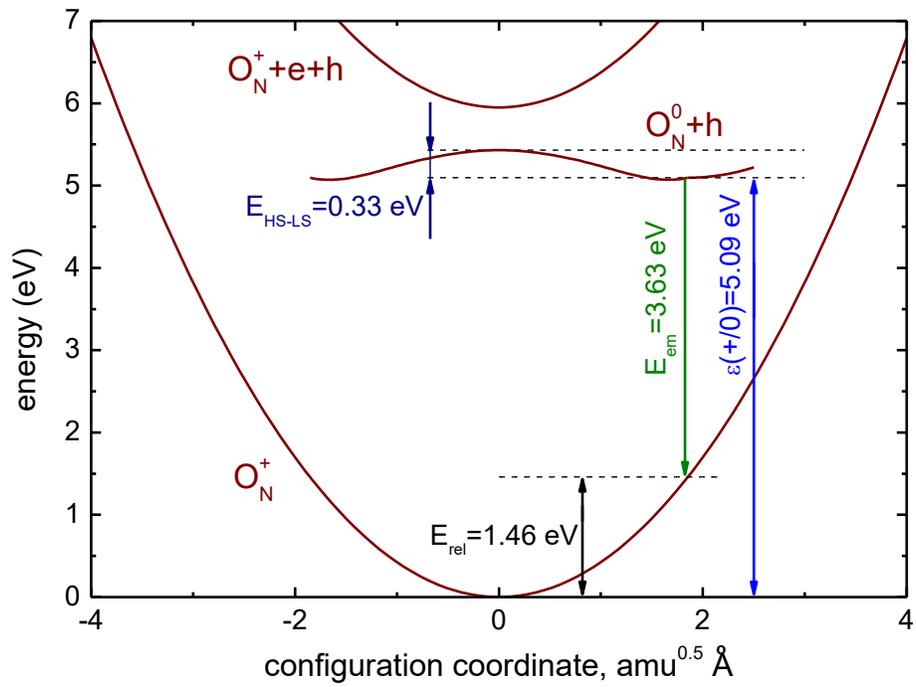

Fig. 3. Configuration coordinate diagram describing hole capture by neutral $O_N$.

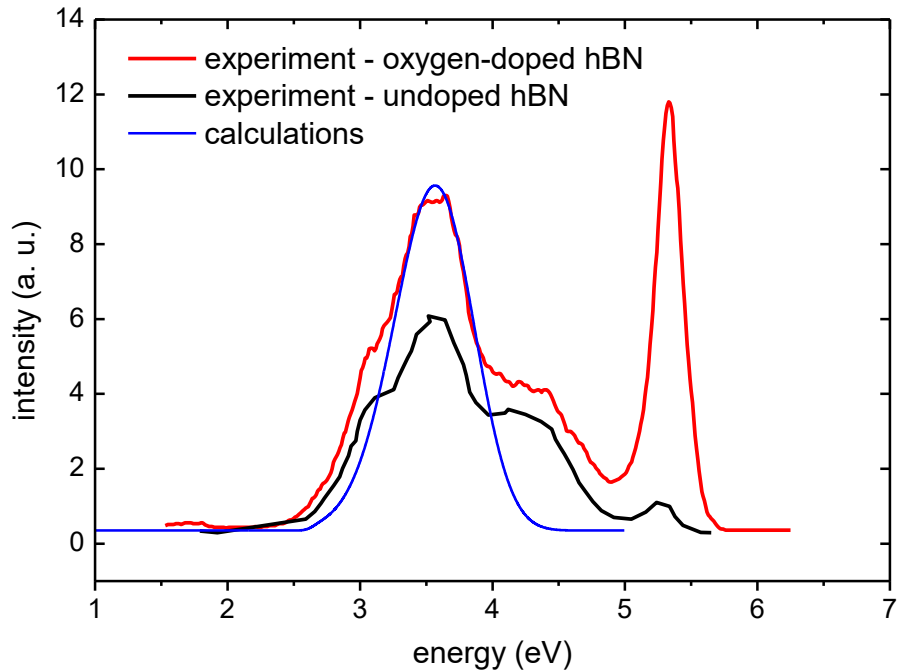

Fig. 4. Calculated luminescence lineshape (blue line) for hole capture by neutral $O_N$, obtained using the single effective-mode (one-dimensional configuration coordinate) approach. For comparison, experimental PL spectra of undoped (black line) and oxygen-doped (red line) hBN from [15] are shown. To facilitate comparison of the lineshapes, the calculated spectrum is rigidly red-shifted by 0.10 eV.